\documentclass[twocolumn,aps,showpacs,nofootinbib,epsfig]{revtex4}
\usepackage{amssymb}
\usepackage{amsmath}
\usepackage{graphics}
\usepackage{epsfig}
\usepackage[dvips]{color}

\newcommand{\be}{\begin{equation}}
\newcommand{\ee}{\end{equation}}
\newcommand{\bea}{\begin{eqnarray}}
\newcommand{\eea}{\end{eqnarray}}
\newcommand{\beas}{\begin{eqnarray*}}
\newcommand{\eeas}{\end{eqnarray*}}

\newcommand{\nn}{\nonumber\\}
\newcommand{\slsh}[1]{{\not \! #1}}

\newcommand{\A}{{\mathbf a}}

\begin{document}
\title{Chiral Symmetry Breaking and Confinement Beyond Rainbow-Ladder Truncation}
\author{Adnan Bashir, Alfredo Raya, Sa\'ul S\'anchez-Madrigal}
\affiliation{Instituto de F\1sica y Matem\'aticas,
Universidad Michoacana de San Nicol\'as de Hidalgo, Edificio C-3,
Ciudad Universitaria, Morelia, Michoac\'an 58040, M\'exico.}

\begin{abstract}

     A non-perturbative construction of the 3-point fermion-boson
vertex which obeys its Ward-Takahashi or Slavnov-Taylor identity,
ensures the massless fermion and boson propagators transform
according to their local gauge covariance relations, reproduces
perturbation theory in the weak coupling regime and provides a
gauge independent description for dynamical chiral symmetry
breaking (DCSB) and confinement has been a long-standing goal in
physically relevant gauge theories such as quantum electrodynamics
(QED) and quantum chromodynamics (QCD). In this paper, we
demonstrate that the same simple and practical form of the vertex
can achieve these objectives not only in 4-dimensional quenched
QED (qQED4) but also in its 3-dimensional counterpart (qQED3).
Employing this convenient form of the vertex \emph{ansatz} into
the Schwinger-Dyson equation (SDE) for the fermion propagator, we
observe that it renders the critical coupling in qQED4
markedly gauge independent in contrast with the bare vertex and
improves on the well-known Curtis-Pennington construction. Furthermore, our proposal
yields gauge independent order parameters for confinement and DCSB
in qQED3.

\end{abstract}


\maketitle

\section{Introduction}\label{I}

Schwinger-Dyson equations (SDEs) are the fundamental field
equations of any quantum field theory. As their derivation
requires no recourse to the value of the interaction strength,
they are ideally suited for perturbative as well as
non-perturbative realms of basic interactions. In particular, they
provide an excellent framework for a unified description of those
field theories for which the evolution of the beta function
encodes diametrically opposed dynamics simultaneously: asymptotic
freedom in the ultraviolet and dynamical chiral symmetry breaking
(DCSB) and confinement in the infrared. Consequently, continuum
studies of the long-range behavior of quantum chromodynamics
(QCD), where effective degrees of freedom are mesons and baryons,
have been vastly carried out through SDEs. It also allows to
extract predictions for the transition region where perturabtive
and non perturbative facets of QCD coexist as well as for the
short distance physics where the fundamental degrees of freedom
are quarks and gluons.

Some original works on DCSB in QCD through SDEs can be dated back
to~\cite{Higa:1984, Miransky:1985}.
 The most natural and practical
truncation of the infinite set of these equations is carried out
at the level of the fermion-boson vertex, see for
example~\cite{Roberts-review:1994, Fischer:2006,
Bashir-Raya-Review:2006, Fischer:1994} for detailed discussions on
the subject. The rainbow-ladder truncation is sufficient to
reproduce a large body of existing experimental data on
pseudoscalar and vector mesons such as their masses, charge radii,
decay constants and scattering lengths as well as their form
factors and the valence quark distribution
functions~\cite{Roberts-Maris:1997, Tandy-Maris:2000,
Tandy-Maris:2002, Ji-Maris:2001, Tandy-Maris:1999,
Tandy-Maris:2003, Maris:2002, Cotanch-Maris:2002,
Bashir-Guerrero-1:2010, Bashir-Guerrero-2:2010,
Tandy-Nguyen:2011}.

However, a fully dressed 3-point fermion-boson vertex is required
to extend the domain of success provided by the SDEs within their
unified description of hadronic physics. For example, although the
$\pi$- and $\rho$-mesons are described rightfully by the
rainbow-ladder truncation, their parity partners, namely, the
$\sigma$- and the $a_1$-mesons are not. The underlying reason has
been recently discovered to be linked with the fact that DCSB
generates a large dressed-quark anomalous chromomagnetic moment.
As a result, spin-orbit splitting between ground-state mesons is
dramatically enhanced. This is the mechanism responsible for a
magnified splitting between parity partners. The essentially
nonperturbative corrections to the rainbow-ladder truncation
largely cancel in the pseudoscalar and vector channels but add
constructively in the scalar and axial-vector channels, providing
a clear signal to go beyond the rainbow-ladder for these
mesons,~\cite{Roberts-Chang:2009, Roberts-Chang:2011}.

On the theoretical side, research efforts spanning a couple of
decades on various gauge field theories such as scalar
QED~\cite{Bashir-Concha:2007, Bashir-Concha:2009,
Bashir-Concha:2010}, spinor QED~\cite{Ball:80, Curtis:90,
Bashir:94, Bashir:96, Kizilersu:95, Kizilersu:09,
Bashir-Guerrero:2011}, field theories in different space-time
dimensions~\cite{Burden:93, Dong:1994, Bashir:98, Bashir:01,
Bashir:04} and QCD~\cite{Skullerud:2003, Skullerud:2007,
Aguilar:2010} have revealed that gauge invariance~\cite{Burden:93,
Bashir:94, Bashir:96, Bashir:thesis, Bashir:02}, gauge covariance
(which is a statement of multiplicative renormalizability of the
2-point function in 4 dimensions)~\cite{Curtis:90,
Bashir-Kizilersu:1998} and  perturbation theory~\cite{Ball:80,
Kizilersu:95, Bashir:99, Davydychev:2000, Bashir:01,
Bashir-Concha:2007} impose severe constraints on the fermion-boson
interaction. The gauge technique of Delbourgo and
Salam~\cite{Salam:1964, Delbourgo-1:1977, Delbourgo-2:1977,
Delbourgo:1979}, introduced decades earlier, was in fact developed
to address some of these constraints, namely the ones which stem
from gauge invariance. This technique culminated in formal results
for the fermion-boson vertex expressed in terms of spectral
functions~\cite{Delbourgo:1981, Delbourgo:1987}. However, this
approach is cumbersome in practical calculations of the fermion
propagator~\cite{Parker:1984, Delbourgo:1984}.

All the studies to-date imply that the 3-point vertex projected
onto the propagator equations is largely determined by the
behavior of the fermion propagator itself and not by the knowledge
of the higher-point functions. There exist numerous {\em
ans}$\ddot{a}${\em tze} for the transverse part of the vertex
(which remains unfixed by the constraints of gauge invariance)
involving different forms of the functional dependence on the
2-point functions, depending upon the case at hand. In this
article, we provide first steps towards a unified approach for
this truncation, applicable to different problems. We employ a
simple and practical form for the full fermion-boson vertex which
respects its Ward-Takahashi identity, yields a fermion propagator
which respects its gauge covariance properties, has the correct
charge conjugation properties and also reproduces its asymptotic
perturbative limit both in QED3 and QED4. Moreover, it not only
renders the critical coupling in qQED4 markedly gauge
independent~\cite{Bashir:thesis} in contrast with the bare vertex
and improves on the Curtis-Pennington vertex~\cite{Curtis:90} but
also yields gauge independent order parameters for confinement and
DCSB in qQED3.


This paper is organized as follows: In Sect.~\ref{II} we decompose
the full fermion-boson vertex into its longitudinal and transverse
parts, invoking Ward-Takahashi identity (WTI) for QED. Employing a
simple {\em ansatz} for the transverse vertex based upon the key
features of qQED4 reduces the gauge dependence of the critical
coupling~\cite{Bashir:thesis} in comparison with even one of the
most sophisticated vertices constructed to date, namely,
Curtis-Pennington vertex~\cite{Curtis:90}. In Sect.~\ref{III}, we
use the same functional form of the {\em ansatz} to study DCSB and
confinement in qQED3. The corresponding order parameters are again
found to be gauge independent. We provide a convincing comparison
with the results obtained by employing the Curtis-Pennington
vertex~\cite{Curtis:90} as well as the Burden-Roberts
vertex~\cite{Burden:1991}. Conclusions and plans for further work
are presented in Sect.~\ref{IV}.

\section{Transverse Vertex: DCSB in QED4}\label{II}

The SDE for the fermion propagator in QED is expressed as
 \bea
 S_F^{-1}(p)&=& S_F^{(0)\,-1}(p)\nn
 &-&4 \pi i \alpha  \int \frac{d^dk}{(2\pi)^d} \ \gamma^\mu S_F(k)
 \Gamma^\nu(k,p) \Delta_{\mu\nu}(q)\;, \label{SDEfpd}
 \eea
in arbitrary space-time dimensions $d$. Here  $\alpha$ is the
electromagnetic coupling (dimensional for $d\neq4$), $S_F(p)$ is
the fermion propagator, $S_F^{(0)}(p)$ is its bare counterpart,
$\Delta_{\mu\nu}(q)$ (with $q=k-p$) is the gauge boson propagator
and $\Gamma^\nu(k,p)$ is the fermion-boson vertex. We can write
the fermion propagator in the following equivalent forms~:
 \bea S_F(p) &=& \sigma_V(p^2)
 \slsh{p}+\sigma_S(p^2) \nn &=&\frac{F(p)}{\slsh{p}-M(p)}\ = \
 \frac{F(p)(\slsh{p}+M(p))}{p^2-M^2(p)} \;,
 \eea
$F(p)$ being the fermion wavefunction renormalization and $M(p)$,
the mass function. Correspondingly, we can write
$S_F^{(0)}(p)=1/(\slsh{p}-m)$, where $m$ denotes the bare fermion
mass. In this article, we work in the chiral limit by setting
$m=0$. In quenched QED, the full gauge boson propagator receives
no radiative corrections, i.e.,
 $$
 \Delta_{\mu\nu}(q)= \Delta_{\mu\nu}^{(0)}(q)=
 -\frac{1}{q^2}\left[g_{\mu\nu}+(\xi-1)\frac{q_\mu q_\nu}{q^2}
 \right]\;,
 $$
where $\xi$ is the usual covariant gauge parameter such that
$\xi=0$ corresponding to the Landau gauge. To be able to solve the
gap equation~(\ref{SDEfpd}), we must know the explicit form of the
fermion-boson interaction $\Gamma^\mu(k,p)$. It is related to the
fermion propagator through the WTI~:
 \be q_\mu
 \Gamma^\mu(k,p) = S_F^{-1}(k)-S_F^{-1}(p)\;.\label{WGTI}
 \ee
 This identity motivates a natural decomposition of the vertex into
 its longitudinal and transverse parts,
 \bea \Gamma^\mu(k,p)=
 \Gamma^\mu_{L}(k,p)+\Gamma^\mu_T(k,p)\;,
 \eea
 where the transverse vertex is defined to be such that
 $q_\mu \Gamma^\mu_T(k,p)=0$ and $\Gamma^{\mu}(p,p)=0$.
 Following Ball and Chiu, we choose the longitudinal part
 of the vertex to be~\cite{Ball:80},
 \bea
 \Gamma^{\mu}_{L}(k,p)&=&\frac{\gamma^{\mu}}{2} \left[
 \frac{1}{F(k)}+\frac{1}{F(p)} \right] \nn &+& \frac{1}{2} \,
 \frac{(\slsh{k}+\slsh{p})(k+p)^{\mu}} {(k^2-p^2)}\left[
 \frac{1}{F(k)}-\frac{1}{F(p)} \right] \nn &-&
 \frac{(k+p)^{\mu}} {(k^2-p^2)}\left[ \frac{M(k)}{F(k)}-
 \frac{M(p)}{F(p)}\right] \;, \nonumber \\
 &\equiv& a(k^2,p^2) \gamma^{\mu} + b(k^2,p^2)
 (\slsh{k}+\slsh{p})(k+p)^{\mu} \nonumber \\
 &+& c(k^2,p^2)
 {(k+p)^{\mu}} \;.
 \label{Lvertex} \eea
 The transverse part is conveniently expressed as~\cite{Ball:80}
 \be
 \Gamma^{\mu}_{T}(k,p)=\sum_{i=1}^{8} \tau_{i}(k^2,p^2,q^2)T^{\mu}_{i}(k,p)
 \;,\label{VT}
 \ee
 where the basis vectors are defined to be~:
\begin{eqnarray}
T^{\mu}_{1}&=&p^{\mu}(k\cdot q)-k^{\mu}(p\cdot q)\nonumber\\
T^{\mu}_{2}&=&\left[p^{\mu}(k\cdot q)-k^{\mu}(p\cdot q)\right]({\not\! k}
+{\not\! p})\nonumber\\
T^{\mu}_{3}&=&q^2\gamma^{\mu}-q^{\mu}{\not \! q}\nonumber\\
T^{\mu}_{4}&=&\left[p^{\mu}(k\cdot q)-k^{\mu}(p\cdot q)\right]k^{\lambda}
p^{\nu}{\sigma_{\lambda\nu}}\nonumber\\
T^{\mu}_{5}&=&q_{\nu}{\sigma^{\nu\mu}}\nn
T^{\mu}_{6}&=&\gamma^{\mu}(p^2-k^2)+(p+k)^{\mu}{\not \! q}\nonumber\\
T^{\mu}_{7}&=&\frac{1}{2}(p^2-k^2)\left[\gamma^{\mu}({\not \! p}+{\not \! k})
-p^{\mu}-k^{\mu}\right]\nn
&&+\left(k+p\right)^{\mu}k^{\lambda}p^{\nu}\sigma_{\lambda\nu}\nonumber\\
 T^{\mu}_{8}&=&-\gamma^{\mu}k^{\nu}p^{\lambda}{\sigma_{\nu\lambda}}
 +k^{\mu}{\not \! p}-p^{\mu}{\not \! k} \eea
 with $
 \sigma_{\mu\nu}=[\gamma_{\mu},\gamma_{\nu}]/2. $
 This special choice of the transverse vertex was put forward by
 Ball and Chiu~\cite{Ball:80}. They carried out a one loop calculation
 of the fermion-boson vertex in the Feynman gauge. They found the
 transverse vertex to be independent of any kinematic singularities
 when $k^2 \rightarrow p^2$. The above choice of the transverse
 basis guarantees that the coefficient of every individual basis vector
 in the Feynman gauge is also free of these singularities .
 It was later pointed out in~\cite{Kizilersu:95} that this attractive feature
 of the basis no longer prevails beyond the Feynman gauge even at the one
 loop level. However, one can re-arrange the basis vectors to restore this
 quality.

 In articles~\cite{Bashir:94,Bashir:96}, Bashir and Pennington proposed
 a family of transverse vertices, which, by construction,
 render the critical value of electromagnetic coupling, above
 which chiral symmetry is restored, completely gauge independent.
 However, the form of the resulting vertex involves intricate
 dependence on the elements which define the fermion propagator. Hence
 its implementation away from the critical coupling is not computationally
 economical. The same is true for the more recent and complete
 construction provided in~\cite{Kizilersu:09} which involves the
 photon momentum $q$ in its construction. However, it is clear from the
 perturbative calculation in~\cite{Davydychev:2000} that an explicit $q^2$
 dependence occurs in every term of each of the $\tau_i$. Therefore, we
 should keep in mind that whenever we neglect the $q^2$ dependence,
 we are only referring to an effective vertex.
 However, there exists an exact relation between the real
 $\tau_i(k^2,p^2,q^2)$ and the effective $\tau_i^{\rm
 eff}(k^2,p^2)$ as spelled out in~\cite{Bashir:98}, and utilized
 in~\cite{Kizilersu:09}. Before we outline this relation, we also
 demand that a
 chirally-symmetric solution should be possible when the bare mass
 is zero, just as in perturbation theory. This is most easily
 accomplished if only those transverse vectors with odd numbers of
 gamma matrices contribute to the transverse vertex. Then the sum in
 Eq.~(\ref{VT}) involves just $i = 2, 3, 6$ and $8$. In the
 chirally symmetric limit, Eq.~(\ref{SDEfpd}) yields~:
\begin{eqnarray}
 \frac{1}{F(p^2)} &=& 1-\frac{e^2}{p^2}\int \frac{d^{d}k}{(2\pi)^{d}}\frac{F(k^2)}{k^2q^2}  \nonumber \\
&& \Big\{a(k^2,p^2)\frac{1}{q^2}[(1-d)q^2(k\cdot p)-2\Delta^2] \nonumber \\
&+& b(k^2,p^2)\frac{1}{q^2}[-2\Delta^2(k^2+p^2)] \nonumber  \\
&-& \frac{\xi}{F(p^2)}\frac{p^2}{q^2}(k^2-k\cdot p) \nonumber \\
&+& \tau_{2}(k^2,p^2,q^2)[-\Delta^2(k^2+p^2)] \nonumber \\
&+& \tau_{3}(k^2,p^2,q^2)[(d-1)q^2(k\cdot p)+2\Delta^2] \nonumber \\
&-& \tau_{6}(k^2,p^2,q^2)[(d-1)(k^2-p^2)k\cdot p] \nonumber \\
&-& \tau_{8}(k^2,p^2,q^2)[\Delta^2(d-2)]   \Big\}  \;,
\label{real}
\end{eqnarray}
where $\Delta^2 = (k \cdot p)^2 - k^2 p^2$. At this stage, it
appears impossible to proceed any further without demanding that
the $\tau_i$ be independent of the angle between the fermion
momentum vectors $k$ and $p$, i.e., independent of $q^2$. This
assumption allows us to carry out the angular integration. In
order to distinguish the transverse components which are assumed
to be independent of $q^2$ from the real ones which explicitly
depend on $q^2$, we can denote the former by $\tau_i^{eff}$. The
equation which then emerges after the angular integration can be
compared to Eq.~(\ref{real}), giving rise to the following exact
relation in arbitrary dimensions~:
 \begin{eqnarray}
  \tau_{2}^{eff}&=& \int d\psi \frac{{\rm sin}^{d-2}\psi}{q^2} \tau_{2} [- \Delta^2] \nonumber \\
  \tau_{3}^{eff}&=&\int d\psi \frac{{\rm sin}^{d-2}\psi}{q^2} \tau_{3}[2 \Delta^2
+ (d-1)(k\cdot p) q^2 ] \nonumber \\
  \tau_{6}^{eff}&=&\int d\psi \frac{{\rm sin}^{d-2}\psi}{q^2} \tau_{6} [k\cdot p] \nonumber \\
  \tau_{8}^{eff}&=&\int d\psi \frac{{\rm sin}^{d-2}\psi}{q^2}
  \tau_{8}[\Delta^2]\;.
 \end{eqnarray}
 For the desired convenience, we have used the compact notation
 $\tau_i^{eff}(k^2,p^2)=\tau_i^{eff}$ and
 $\tau_i(k^2,p^2,q^2)=\tau_i$.
 This relationship, of course, depends upon
 the space-time dimension $d$. It allows us to propose an
 {\em ansatz} for an
 effective but simple $q^2-$independent vertex which fulfills the
 general requirements that any transverse vertex must satisfy~:
 \bea
 \Gamma^{\mu}_T(k,p) &=&
 \sum_{i=2,3,6,8} \tau_{i}^{\rm eff}(k^2,p^2)T^{\mu}_{i}(k,p) \;,
 \eea
where
 \bea
 \nn \tau_2^{\rm eff}(k^2,p^2) &=&
 \frac{\A_2^d}{(k^4-p^4)} \;
 \left[\frac{1}{F(k)} -\frac{1}{F(p)}\right]  \;,  \\
 \nn \tau_3^{\rm eff}(k^2,p^2) &=&  \frac{\A_3^d}{(k^2-p^2)} \;
 \left[\frac{1}{F(k)} -\frac{1}{F(p)}\right]  \;, \\
 \nn \tau_6^{\rm eff}(k^2,p^2) &=&
 \frac{\A_6^d(k^2+p^2)}{(k^2-p^2)^2} \;
 \left[\frac{1}{F(k)} -\frac{1}{F(p)}\right]  \;,  \\
 \nn \tau_8^{\rm eff}(k^2,p^2) &=&  \frac{{\mathbf a}_8^d}{(k^2-p^2)}
 \; \left[\frac{1}{F(k)} -\frac{1}{F(p)}\right]  \;.
 \eea
 This construction draws on a direct comparison with the structural
 dependence of the longitudinal vertex on the elements which make
 up the fermion propagator. Special care has been taken such that
 the perturbative limit of the transverse vertex conforms with
 its one loop expansion in the asymptotic limit of $k^2 >> p^2$.
 Moreover, it is required to transform correctly under the charge
 conjugation and parity operations.

 Due to the dimension-dependence of the exact connection of these effective
 $\tau_i(k^2,p^2)$ with the real $\tau_i(k^2,p^2,q^2)$, the least we expect
 is that the coefficients $\A_i^d$ would depend on the space-time
 dimensions, justifying the use of the symbol.
In 4 space-time dimensions, parameters $\A_i$ are constrained by
the requirement of multiplicative renormalizability of the
massless fermion propagator in the following manner~:
 \bea
 1+\A_2^4+2\A_3^4+2\A_8^4-2\A_6^4&=&0 \;.
 \label{MR}
 \eea
Additionally, one loop perturbative calculation of the transverse
fermion-boson vertex in an arbitrary covariant gauge reveals that
 \bea
 \Gamma^{\mu}_{T}(k,p) &\stackrel{k^2 >> p^2}{=}& - \frac{\alpha \xi}{8 \pi} \; {\rm ln}
 \frac{k^2}{p^2} \; \left[ \gamma^{\mu} - \frac{k^{\mu} {\not \! k}}{k^2} \right]
 \eea
This perturbative condition imposes the following constraint on
the $\A_i$~:
 \bea
 \A_3^4+\A_6^4 &=& \frac{1}{2} \;.
 \eea
It is worth noting that the choice $\A_6^4=1/2$, $\A_3^4=0$
corresponds to the Curtis-Pennington vertex~\cite{Curtis:90}.
Enjoying a broader choice of available parameters, which also
includes $\A_2^4$ and $\A_8^4$ (taken to be zero
in~\cite{Curtis:90}), we expect to construct an improved
truncation of the SDEs. It is easy to verify that with the choice of
the transverse vertex defined by
 \be \A_6^4=-\frac{1}{2}\qquad
 \mbox{and} \qquad \A_2^4=\frac{11}{4} \;\label{BP4d}
 \ee
and then inserted into the gap equation~(\ref{SDEfpd}),
that the critical value of the coupling for masses to be
dynamically generated, i.e., $\alpha_c$, turns out to be
considerably more gauge independent for a broad range of values of
the covariant gauge parameter $\xi$ not only as compared to the
bare vertex but also to the Curtis-Pennington vertex by a fair
amount of margin~\cite{Curtis:90, Bashir:thesis}. This has been
depicted in~Fig.~(\ref{fig:alphac}). We now turn our attention to
QED3.

\begin{figure}[t]
{\centering {\epsfig{file=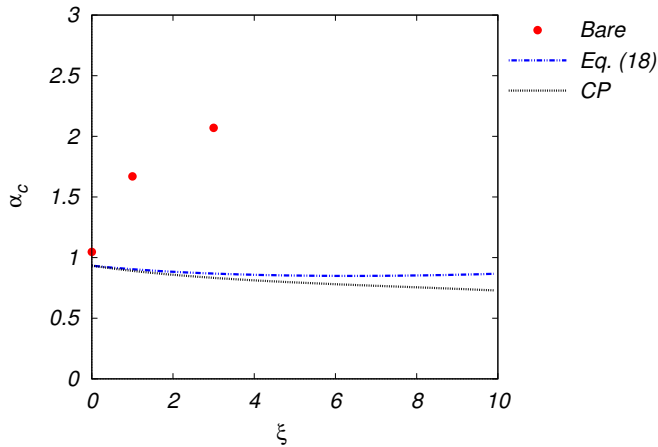,
width=0.7\columnwidth,angle=270}}
\par}
\caption{Critical coupling in QED4 as a function of the gauge
parameter. The choice of the vertex suggested in Eq.~(\ref{BP4d})
renders the critical coupling in qQED4 markedly gauge independent
in contrast with the bare vertex and also improves on the
Curtis-Pennington (CP) vertex. } \label{fig:alphac}
\end{figure}

\begin{figure}[t]
{\centering {\epsfig{file=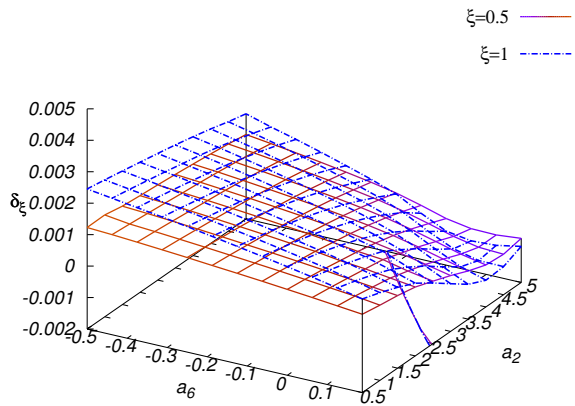,
width=0.8\columnwidth,angle=270}}
\par}
\caption{Gauge difference of the condensate as compared to the
Landau gauge, Eq.~(\ref{diff}), over the $\A_2/\A_6$-plane. Solid
surface corresponds to $\xi=0.05$, dot-dashed surface to $\xi=1$
and the dotted surface to $\xi=1.5$. Contour of gauge independence
of the condensate is along the straight line given by
Eq.~(\ref{line}). The scale of the height of the surfaces is set
by the value $e^2=1$.} \label{fig:condgauge}
\end{figure}

\section{Gauge Independence in QED3}\label{III}

 Quantum electrodynamics in (2+1)-dimensions, i.e. QED3, is an
interesting theory. It exhibits confinement and DCSB. Therefore,
for the last three decades, it has served as a toy model for QCD
to deepen our understanding of these fascinating yet complicated
phenomena through the efficient tools of SDEs as well as
lattice~\cite{Pisarski:1984, Appelquist:1985, Pennington:1991,
Pennington:1992, Strouthos:2002, Strouthos:2004,
Bashir-Roberts:2008, Zong:2008, Fischer:2009, Zong:2010,
Strouthos:2010, Swanson:2011, Fischer:2011}. It is also of
interest in condensed matter physics as an effective field theory
for high-temperature
superconductors~\cite{Franz:01,Franz:02,Tesanovic:02,Herbut:02,Thomas:07}
and graphene~\cite{Novoselov:05,Gusynin:07, Riazuddin:2011}.

In all gauge theories including QED3, among the covariant gauges,
Landau gauge occupies a special place for a number of theoretical
reasons: wavefunction renormalization receives no contribution at
the one loop level in any space-time
dimensions~\cite{Bashir:04}\footnote{This is one reason why
$\gamma^\mu$ is a good choice for the vertex in this particular
gauge.}, it is a fixed-point of the  renormalization group and it
is the gauge in which the infrared behavior of the fermion
propagator is neither enhanced nor suppressed by a non-dynamical
gauge-dependent exponential factor arising from a gauge
transformation, as dictated by the Landau-Khalatnikov-Fradkin
transformations
(LKFT)~\cite{Landau:55,Fradkin:55,Johnson:59,Zumino:60}.
Therefore, one stands the best chance to provide a reliable {\em
ansatz} of the fermion-boson vertex in this gauge than in any
other. Results can then be simply translated to other gauges by
means of the LKFT. Such a strategy is a bit involved to implement
for $d=4$. However, it has successfully been applied in QED3 in
Refs.~\cite{Burden:92, Bashir:07, Bashir:00b, Bashir:02b,
Bashir:05, Bashir:09, Madrigal:2008}.

\begin{figure}[t]
{\centering {\epsfig{file=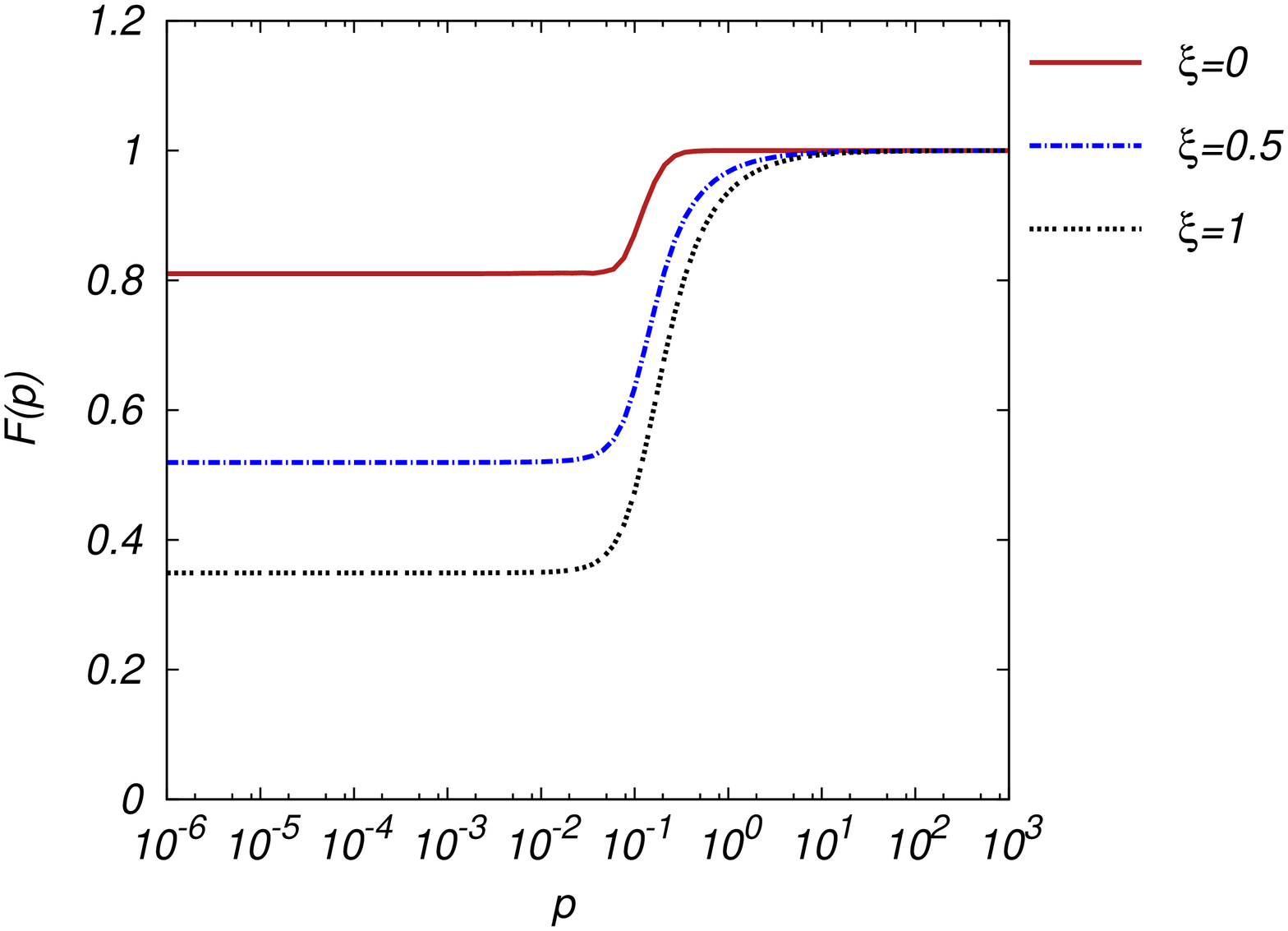,
width=0.7\columnwidth,angle=270} \epsfig{file=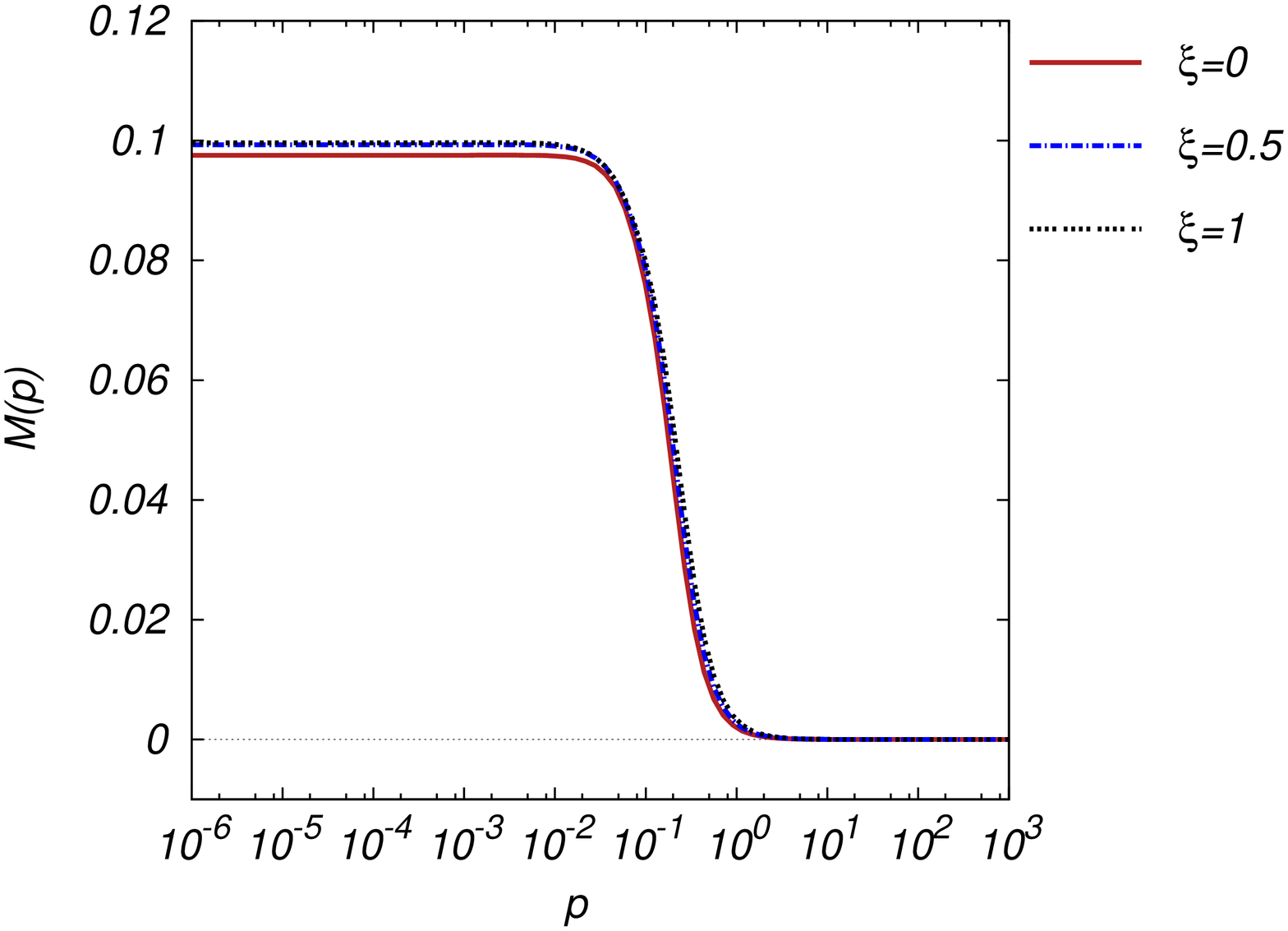,
width=0.7\columnwidth,angle=270}}
\par}
\caption{Fermion propagator in different covariant gauges. Solid
line corresponds to $\xi=0$, dot-dashed line to $\xi=0.5$ and
dotted line to $\xi=1$. {\it Upper panel:} Wavefunction
renormalization $F(p)$. {\it Lower panel:} Mass function $M(p)$.
The scale of the curves is set by the value $e^2=1$.}
\label{fig:propvertex}
\end{figure}

However, the fact remains that an LKFT for the fermion propagator
as well as the 3-point vertex itself must conspire in such a
manner as to yield a full 3-point vertex which would render
physical observables independent of the gauge parameter, no matter
what gauge we choose to work with. Precisely with this idea in
mind, we presented the construction of the vertex in QED$_4$ in
the previous section. We now ask ourselves whether the same form
of the vertex would be sufficient to implement gauge invariance in
three space-time dimensions. This implies finding $\A_i$ in $d=3$
dimensions. For $d=3$, one loop perturbative calculation of the
transverse fermion-boson vertex in an arbitrary covariant gauge
comes out to be
 \bea
 \Gamma^{\mu}_{T}(k,p) &\stackrel{k^2 >> p^2}{=}& - \frac{\alpha \xi \pi}{8 p} \;
 \left[ \gamma^{\mu} - \frac{k^{\mu} {\not \! k}}{k^2} \right] \;.
 \eea
Interestingly, just as for $d=4$, we again have~:
 \bea
 \A_3^3+\A_6^3 &=& \frac{1}{2}
 \eea
in 3-dimensions. Similarly, the condition for the multiplicative
renormalizability translates as the one of LKFT for the fermion
propagator. Therefore, we now proceed to solve the gap equation in
QED3 by employing the same form of the vertex as for QED4. Our
starting point is to explore the configuration space of $\A_2$ and
$\A_6$ in the Landau gauge. Then, we change the gauge parameter
infinitesimally and repeat the same exercise. We look for the
domain in the $\A_2/\A_6$-plane for which the difference of the
condensate,
 \be
 \delta_\xi=\langle\bar\psi\psi\rangle_\xi-\langle\bar\psi\psi\rangle_{\xi=0}\;,\label{diff}
 \ee
is minimal. This is illustrated in Fig.~\ref{fig:condgauge}.
Within our numerical accuracy, different surfaces for fixed $\xi$
intersect along a line parameterized by
 \bea
 \A_2^3 &=& 3.81-7.6\ \A_6^3\;.\label{line}
 \eea
Thus, there is a family of vertex \emph{ans\"atze} which yield a
gauge invariant value of the condensate! Below we carry out a
study of DCSB and confinement by selecting one member of this
family of vertices. {\em A priori}, there is no guarantee that
gauge independent DCSB should imply gauge independent confinement
or vice versa.

\subsection{DCSB}

In order to make contact with our studies in QED4, we select the
value of $\A_2^3=2.75$ and define our vertex by fixing $\A_6$
according to Eq.~(\ref{line}). With the choice of this vertex, we
solve the gap equation in different gauges. Results are shown in
Fig~\ref{fig:propvertex}.

The mass functions change only slightly, whereas the variation in
$F(p)$ is more noticeable. These changes conspire with each other
to render the chiral fermion condensate gauge independent. We
carry out a comparison with the same quantity obtained from the
Curtis-Pennington~\cite{Curtis:90} vertex as well as the
Burden-Roberts vertex~\cite{Burden:1991}. The results have been
plotted in~Fig~.\ref{fig:d3gauge}, which clearly demonstrate the
superiority of our proposal over the previous similar efforts for
the de-construction of this Green function.

\begin{figure}[t]
{\centering {\epsfig{file=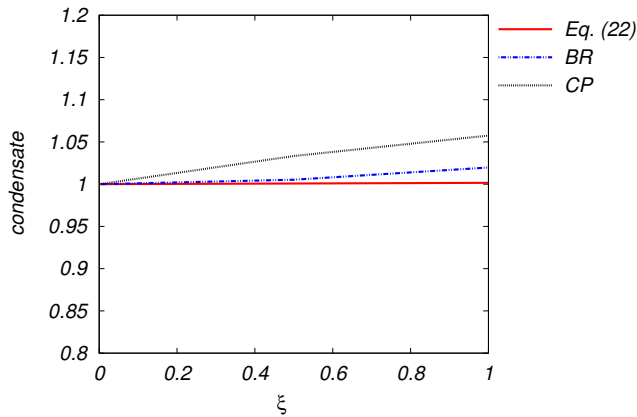,
width=0.65\columnwidth,angle=270}}
\par}
\caption{Gauge dependence of the chiral quark condensate for QED3.
We compare our proposal against the ones by Burden-Roberts as well
as Curtis-Pennington.} \label{fig:d3gauge}
\end{figure}

\begin{figure}[t]
{\centering {\epsfig{file=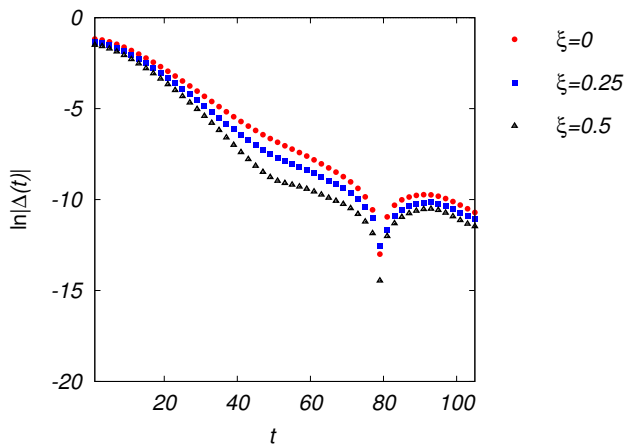,
width=0.7\columnwidth,angle=270}}
\par}
\caption{Spatially averaged Schwinger function in different
gauges. Solid pentagons correspond to $\xi=0$, solid diamonds to
$\xi=0.25$ and solid squares to $\xi=0.5$. Oscillations are
inferred from the periodic peaks, which signal confinement.
Position of the first dip is an order parameter for confinement.
It is gauge invariant for our construction of the interaction.}
\label{fig:confosc}
\end{figure}

\subsection{Confinement}

Confinement can be realized through the violation of the axiom of
reflection positivity. For the fermion propagator, breach of the
said axiom entails that the elementary excitation described by
$S_F(p)$ cannot appear in the Hilbert space of observables.
Confinement in QED3 can be explored through the positivity of the
spatially averaged Schwinger function\footnote{An alternative test
was performed in Refs.~\cite{Hofmann:10,Hofmann:11} for the vector
part of the propagator.} \be \Delta(t) = \int
d^2x\int\frac{d^3p}{(2\pi)^3}e^{ip\cdot x} \sigma_s(p^2)\;, \ee
which we construct from the solutions shown in
Fig.~\ref{fig:propvertex}. In Fig.~\ref{fig:confosc} we display
the logarithm of the absolute value of $\Delta(t)$ in different
gauges. An oscillatory behavior of this function is revealed by
the pronounced periodic peaks. This implies that $\Delta(t)$ is
not positive definite and thus confinement is realized. The
corresponding propagator possesses a pair of complex conjugated
mass poles~\cite{Krein:92}. Denoting $t_c$ the position of the
first oscillation, $\nu=1/t_c$ serves as an order parameter for
confinement~\cite{Hawes:94, Bashir:05, Bashir:09}. Noticeably,
$\nu$ is the same in all gauges, within our numerical accuracy.
Thus confinement too has come out to be gauge independent with our
choice of the fermion-boson vertex.

\begin{figure}[t]
{\centering {\epsfig{file=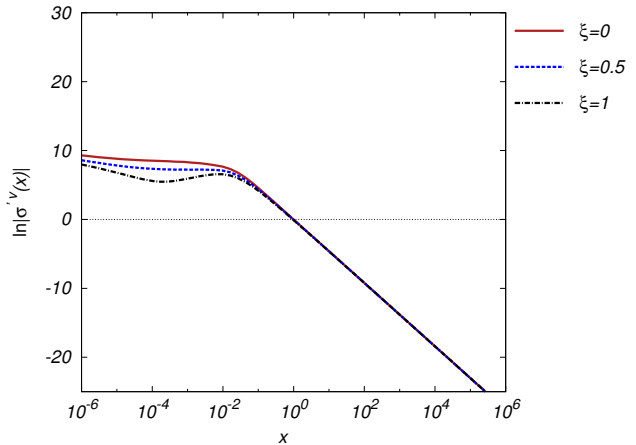,
width=0.7\columnwidth,angle=270}}
\par}
\caption{The first derivative of $\sigma_v(x)$ in various gauges.
The maximum developed by all these functions is the order
parameter, which is the same in all gauges within our numerical
accuracy.} \label{fig:confder}
\end{figure}

An alternative view of confinement stems from the fact that any
function with an inflection point must violate the axiom of
reflection positivity~\cite{Bashir:09, Hofmann:10, Hofmann:11}.
Let $x=p^2$. The above statement then implies that if there exists
a point $x_c>0$ such that
 \be
  \left.
 \frac{d^2}{dx^2}\sigma_v(x)\right|_{x=x_c}=0\;,
 \ee
then the propagator describes a confined excitation, and $x_c$
plays the role of an order parameter for confinement. In
Fig.~\ref{fig:confder}, we plot the logarithm of the first
derivative of $\sigma_v(x)$ in different gauges. We observe that
all these curves develop a maximum at the same point $x_c$, which
again means that there is confinement, and the corresponding order
parameter is independent of choice of the covariant gauge $\xi$.

\section{Concluding Remarks}~\label{IV}

In this article, building upon the proposal put forward
in~\cite{Bashir:thesis}, we employ a simple form (e.g., it is
independent of the photon momentum $q$) for the fermion-boson
vertex in arbitrary dimensions. In 4-dimensions, it ensures WTI is
satisfied, massless fermion propagator is multiplicatively
renormalizable, one loop perturbation theory is recovered in the
asymptotic limit $k^2 >> p^2$, charge conjugation and parity
properties of the vertex are respected and gauge independent value
of critical electromagnetic coupling is achieved below which
chiral symmetry is restored. This construction involves two free
parameters which are momentum and gauge independent. However, the
price we pay for ignoring the photon momentum dependence in this
vertex {\em ansatz} is that these parameters naturally depend upon
the dimension of space-time we choose to work with. We demonstrate
that the same simple form of the vertex is able to render the
order parameters for DCSB and confinement gauge invariant also in
qQED3. We provide explicit comparisons with earlier proposals such
as the Curtis-Pennington vertex (designed for $d=4$) as well as
the Burden-Roberts vertex (constructed for $d=3$) to bring out the
fact that our construction offers a marked improvement.

This is a first step in our intent to provide a unified truncation
scheme for different gauge theories and a large body of associated
physical observables. A natural next step is to extend our ideas
to the SDE study of QCD. As mentioned before, an improved
understanding of hadronic masses invokes additional structures in
the fermion-boson vertex involving anomalous electromagnetic and
chromomagnetic moments for dynamically massive quarks in the
infrared. However, as a word of caution, one should remember that
QCD is markedly more involved than QED3 as well as QED4. In
covariant gauge QCD, ghosts play a vital role for its infrared
dynamics, at least in the Landau gauge. Naturally, the ghost-gluon
interaction also enters into the picture and one has to take into
account the resulting complications with appropriate care. This work
is in progress. \\

\noindent \acknowledgements{AB and AR are grateful for the
financial support from SNI, CONACyT and CIC-UMSNH grants. SSM
acknowledges CONACyT funding for his postgraduate studies. We
thank Roc\1o Berm\'udez for generating some numerical data for us.}

\end{document}